\documentstyle[12pt]{article}

\makeatletter
\@addtoreset{equation}{section}
\makeatother

\def\vep{\varepsilon}

\def\K{{\cal K}}
\def\W{{\cal W}}
\def\T{{\cal T}}

\def\pa{{\parallel}}
\def\pe{{\perp}}

\def\be{\begin{equation}}
\def\ee{\end{equation}}
\def\bea{\begin{eqnarray}}
\def\eea{\end{eqnarray}}

\begin{document}

\begin{flushright}
\vspace{1mm}
 FIAN/TD/07--03\\
\vspace{-1mm}
\end{flushright}\vspace{1cm}

\begin{center}
{\large\bf Nonlinear Equations for Symmetric
Massless Higher Spin Fields in $(A)dS_d$}
\vglue 0.6  true cm
\vskip0.8cm
{M.A.~Vasiliev}
\vglue 0.3  true cm

I.E.Tamm Department of Theoretical Physics, Lebedev Physical Institute,\\
Leninsky prospect 53, 119991, Moscow, Russia
\vskip1.3cm
\end{center}

\begin{abstract}
Nonlinear field equations for totally symmetric
bosonic massless fields of all spins  in any dimension are presented.
\end{abstract}

\newcommand{\ty}{\hat{y}}
\newcommand{\bee}{\begin{eqnarray}}
\newcommand{\eee}{\end{eqnarray}}
\newcommand{\nn}{\nonumber}
\newcommand{\lis}{Fort1,FV1,LV}
\newcommand{\hy}{\hat{y}}
\newcommand{\by}{\bar{y}}
\newcommand{\bz}{\bar{z}}
\newcommand{\go}{\omega}
\newcommand{\e}{\epsilon}
\newcommand{\f}{\frac}
\newcommand{\p}{\partial}
\newcommand{\half}{\frac{1}{2}}
\newcommand{\ga}{\alpha}
\newcommand{\gal}{\alpha}
\newcommand{\U}{\Upsilon}
\newcommand{\C}{{\bf C}}
\newcommand{\ups}{\upsilon}
\newcommand{\bu}{\bar{\upsilon}}
\newcommand{\dga}{{\dot{\alpha}}}
\newcommand{\dgb}{{\dot{\beta}}}
\newcommand{\gb}{\beta}
\newcommand{\gga}{\gamma}
\newcommand{\gd}{\delta}
\newcommand{\gl}{\lambda}
\newcommand{\gk}{\kappa}
\newcommand{\gep}{\epsilon}
\newcommand{\gvep}{\varepsilon}
\newcommand{\gs}{\sigma}
\newcommand{\V}{|0\rangle}
\newcommand{\ws}{\wedge\star\,}
\newcommand{\gee}{\epsilon}
\newcommand{\ggg}{\gamma}
\newcommand\ul{\underline}
\newcommand\un{{\underline{n}}}
\newcommand\ull{{\underline{l}}}
\newcommand\um{{\underline{m}}}
\newcommand\ur{{\underline{r}}}
\newcommand\us{{\underline{s}}}
\newcommand\up{{\underline{p}}}

\newcommand\uq{{\underline{q}}}
\newcommand\ri{{\cal R}}
\newcommand\punc{\multiput(134,25)(15,0){5}{\line(1,0){3}}}
\newcommand\runc{\multiput(149,40)(15,0){4}{\line(1,0){3}}}
\newcommand\tunc{\multiput(164,55)(15,0){3}{\line(1,0){3}}}
\newcommand\yunc{\multiput(179,70)(15,0){2}{\line(1,0){3}}}
\newcommand\uunc{\multiput(194,85)(15,0){1}{\line(1,0){3}}}
\newcommand\aunc{\multiput(-75,15)(0,15){1}{\line(0,1){3}}}
\newcommand\sunc{\multiput(-60,15)(0,15){2}{\line(0,1){3}}}
\newcommand\dunc{\multiput(-45,15)(0,15){3}{\line(0,1){3}}}
\newcommand\func{\multiput(-30,15)(0,15){4}{\line(0,1){3}}}
\newcommand\gunc{\multiput(-15,15)(0,15){5}{\line(0,1){3}}}
\newcommand\hunc{\multiput(0,15)(0,15){6}{\line(0,1){3}}}
\newcommand\ls{\!\!\!\!\!\!\!}

\section{Introduction}

Higher spin (HS) gauge theories are  theories of gauge fields of all
spins (see e.g. \cite{gol} for  a review).
 Because HS gauge symmetries are infinite-dimensional,
HS gauge theories may correspond to
most symmetric vacua of a theory of fundamental interactions
presently identified with superstring theory.
The problem is to introduce interactions of  HS fields
in a way compatible with  nonabelian
HS gauge symmetries containing diffeomorphisms and Yang-Mills symmetries.
Full nonlinear dynamics of HS gauge fields
has been  elaborated so far at the level of equations of motion
for $d=4$ \cite{4d,more} which is the simplest nontrivial case
since HS gauge fields do not propagate if $d<4$.
Some lower order interactions of HS fields in the framework
of gravity were
worked out at the action level for $d=4$  \cite{FV1}
and for $d=5$  \cite{5d}. As a result, it was found out that
(i) consistent HS theories contain infinite sets of
infinitely increasing spins;
(ii) HS gauge
interactions contain higher derivatives;
(iii) in the framework of gravity, unbroken
HS gauge symmetries require a non-zero cosmological constant;
(iv)  HS symmetry algebras
\cite{FVa} are certain star product algebras
\cite{Fort2}.
The properties (i) and (ii) were deduced in the
earlier work  \cite{pos}
on  HS interactions in flat space.

The feature that unbroken
HS gauge symmetries require a non-zero cosmological constant
is  crucial in several respects. It
explains why the analysis of HS--gravitational interactions
in the framework of the expansion near the flat
background led to negative conclusions  \cite{diff}.
The $S$-matrix
Coleman-Mandula-type no-go arguments \cite{cm} become
irrelevant because there is no $S$-matrix in the $AdS$ space.
Also it explains why the HS gauge theory phase is not
directly observed in the  superstring theory  prior
its full formulation in the $AdS$ background is found, and
 fits the idea of the correspondence between
HS gauge theories in the bulk and the boundary conformal theories
\cite{Sun,Wit,3d,SS,BHS,KP}.

In the context of  applications of the
HS gauge theory to the superstring theory ($d=10$) and
M theory ($d=11$), it is important  to extend
the $4d$ results on the HS--gravitational interactions
to higher dimensions. The aim of this paper is to present
the full nonlinear formulation of the field equations for
totally symmetric bosonic HS fields in any dimension.
The form of the proposed equations  is analogous to that of
the $4d$ equations of \cite{more}.

\section{Free higher spin gauge fields}
\label{Free Fields}
There are two equivalent approaches to description of
totally symmetric bosonic massless fields of all spins at the free
field level.
The approach developed by Fronsdal \cite{Fr} and de Wit and
Freedman \cite{WF} is parallel to the metric formulation of
gravity.  Here an
integer spin $s$ massless field is described by a totally
symmetric tensor $\varphi_{a_1\ldots a_s}$ ($a,b,c = 0\ldots d-1$)
subject to the double
tracelessness condition \cite{Fr}
$\varphi^{b}{}_{b}{}^{c}{}_{c a_5\ldots a_s}=0$ which is nontrivial
for $s\geq 4$. The free field Abelian HS gauge transformation is
\begin{equation}
\delta\varphi_{a_1\ldots a_s}=\partial_{\{a_1}
\varepsilon_{a_2\ldots a_{s}\}}\,,
\end{equation}
where the parameter $\varepsilon_{a _1...a_{s-1}}$ is
a totally symmetric rank $(s-1)$ traceless tensor,
$
\varepsilon^{b}{}_{b a_3\ldots a_{s-1}}=0\,.
$
A nonlocal version of the same theory with unrestricted
gauge parameters was developed in \cite{FS}.

An alternative approach operates in terms of a 1-form frame like
HS gauge field $e_{a_1\ldots a_{s-1}}=dx^\un e_{\un\,a_1\ldots a_{s-1}}$
\cite{GHS,Fort1,LV} which is traceless in the  fiber indices,
\be
\label{etr}
e_b{}^{ba_3\ldots a_{s-1}}=0\,.
\ee
(We use  underlined letters for  indices of vector fields and forms.)
The  Abelian HS gauge transformation law in the flat space-time is
\be
\label{ltr}
\delta e^{a_1\ldots a_{s-1}} = d \vep^{a_1\ldots a_{s-1}}
+ h_b\vep^{a_1\ldots a_{s-1}\,,b}\,.
\ee
Here $h^a$ is the flat
space frame field (i.e., $h^a = dx^a$ in the Cartesian coordinates).
$\vep^{a_1\ldots a_{s-1}}$ is a totally symmetric traceless
gauge parameter equivalent to that of the Fronsdal's formulation.
The gauge parameter
$\vep^{a_1\ldots a_{s-1}\,,b}$ is also traceless and satisfies the
condition that the symmetrization over all its indices is zero
$\vep^{\{a_1\ldots a_{s-1}\,,a_s\}}=0$. It is a HS generalization of
the parameter of the local Lorentz transformations in gravity
($s=2$) which is  $\vep^{a\,,b}$.  The
Lorentz type ambiguity due to $\vep^{a_1\ldots a_{s-1}\,,b}$
can be gauge fixed by requiring the
frame type HS gauge field to be totally symmetric
\be
\label{gfix}
e_{n\, a_2\ldots a_{s}}=\varphi_{n a_2\ldots a_s}\,,
\ee
thus establishing equivalence with the Fronsdal's formulation.
Note that such defined $\varphi_{ a_1\ldots a_s}$ is automatically
double traceless as a consequence of (\ref{etr}).

The Lorentz type HS symmetry with the parameter
$\vep^{a_1\ldots a_{s-1}\,,b}$ assumes a HS 1-form
connection $\go^{a_1\ldots a_{s-1}\,,b}$. {}From the analysis of
its transformation law it follows \cite{Fort1,LV} that, generically,
some new gauge connections and symmetry parameters have to be introduced.
As a result,
the full set of HS connections associated with a spin $s$ massless
field consists of the 1-forms
$dx^\un \go_\un{}^{a_1 \ldots a_{s-1}, b_1\ldots b_t }$
which take values in  all irreducible
representations of the $d$-dimensional massless Lorentz group
$o(d-1,1)$ described by the traceless Young tableaux with at most
two rows such that the upper row has length $s-1$
\bigskip
\vskip -5mm
\be
\label{dia}
\begin{picture}(20,50)
\put(20,45){s-1}
\put(33,35){\circle*{2}}
\put(25,35){\circle*{2}}
\put(17,35){\circle*{2}}
\put(25,25){\circle*{2}}
\put(17,25){\circle*{2}}
\put(33,25){\circle*{2}}
\put(00,40){\line(1,0){70}}
\put(00,30){\line(1,0){70}}
\put(50,30){\line(0,1){10}}
\put(60,30){\line(0,1){10}}
\put(70,30){\line(0,1){10}}
\put(00,20){\line(1,0){50}}
\put(00,20){\line(0,1){20}}
\put(10,20){\line(0,1){20}}
\put(40,20){\line(0,1){20}}
\put(50,20){\line(0,1){20}}
\put(20,10){t}
\end{picture}
\ee
In other words, the  1-forms
$dx^\un \go_\un {}^{a_1 \ldots a_{s-1}, b_1\ldots b_t }$  are
symmetric in the Lorentz vector indices
$a_i$ and $b_j$ separately and satisfy the relations
$
\go_\un{}^{\{a_1 \ldots a_{s-1}, a_s\} b_2\ldots b_t }   =0
$,
$
\go_\un{}^{a_1 \ldots a_{s-3}c}{}_c{}^{, b_1\ldots b_t } =0\,.
$
({}From here it follows
that all other traces of the fiber indices are also zero.)
For the case  $t=0$, the field $\go^{a_1 \ldots a_{s-1}}$
identifies with the dynamical spin $s$ frame type field
$e^{a_1\ldots a_{s-1}}$.

The formalism of \cite{Fort1,LV} works both in
$(A)dS_d$ and in flat space. In $(A)dS_d$
it allows to build the free
HS actions in terms of manifestly gauge invariant linearized
(Abelian) field strengths. Explicit form of these linearized
curvatures provides the starting point towards determination
of nonabelian HS symmetries and HS curvatures.
For the case of $d=4$ this program was realized in
\cite{Fort1,FVa}. Here we extend these results to the bosonic HS theory
in any dimension. To this end it is  convenient to use the
observation of \cite{5d} that the
 collection of the HS  1-forms
$ \go{}^{a_1 \ldots a_{s-1}, b_1\ldots b_t }$
with all $0\leq t \leq s-1$ can be
interpreted as a result of the ``dimensional reduction''
of a 1-form $ \go{}^{A_1 \ldots A_{s-1}, B_1\ldots B_{s-1} }$
carrying the irreducible representation of  $o(d-1,2)$ or $o(d,1)$
$ (A,B = 0,\ldots , d)$ described by the traceless two-row
rectangular Young tableau of length $s-1$
\be
\label{irre}
\go^{\{A_1 \ldots A_{s-1},A_s\} B_2\ldots B_{s-1} } =0\,,\qquad
\go^{A_1 \ldots A_{s-3}C}{}_{C,}{}^{B_1\ldots B_{s-1} } =0\,.
\ee

Let us first recall how this approach works in the gravity case.
$d$ dimensional gravity can be described
 by a 1-form connection  $\go^{AB}= - \go^{BA}$
of the $(A)dS$ Lie algebra $(o(d,1))$  $o(d-1,2)$.
The Lorentz subalgebra
$o(d-1,1)$  is identified  with the
stability subalgebra of some vector $V^A$. Since
we are discussing local Lorentz symmetry, this vector can be chosen
differently in different points of space-time, thus becoming a
field  $V^A = V^A (x)$. The norm of this vector is convenient
to relate to the cosmological constant
$\Lambda$ so that $V^A$ has dimension of length
\be
\label{vnorm}
V^AV_A = -\Lambda^{-1}\,.
\ee
$\Lambda$ is supposed to be negative and positive in
 the $AdS$ and $dS$ cases, respectively (within the mostly minus
signature). This allows for a covariant definition of the
frame field and
Lorentz connection \cite{SW,PrV}
\be
\label{defh}
 E^A = D(V^A) \equiv d V^A + \go^{AB}V_B\,,\qquad
\go^{L\,AB} = \go^{AB} +\Lambda  ( E^A V^B - E^B V^A )\,.
\ee
According to these definitions
$
E^A V_A =0\,,
$
$
D^L V^A = dV^A + \go^{L\,AB}V_B \equiv 0\,.
$
When the  frame $E_\un^A$ has the maximal rank $d$ it
gives rise to a nondegenerate metric tensor
$
g_{\un\um} = E_\un^A E_\um^B \eta_{AB}\,.
$
The torsion 2-form is
$
r^A\equiv DE^A\equiv  r^{AB} V_B \,.
$
The zero-torsion condition
$
r^A = 0\,
$
expresses the Lorentz connection via
(derivatives of) the frame field in a usual manner. The
$V^A$ transversal components of the curvature (\ref{Rvac}) $r^{AB}$
identify with the Riemann tensor shifted by the term  bilinear in the
frame 1-form. As a result, any field $\go_0$ satisfying
the zero-curvature equation
\be
\label{Rvac}
r^{AB}= d\go_0^{AB} + \go_0^{A}{}_C \wedge \go_0^{CB}=0\,,
\ee
describes $(A)dS_d$ space-time
with the cosmological term $\Lambda$ provided that the metric
tensor is nondegenerate.

The Lorentz irreducible HS connections
$ \go{}^{a_1 \ldots a_{s-1}, b_1\ldots b_t }$
are the $d$-dimensional traceless parts of
those components of
$ \go{}^{A_1 \ldots A_{s-1}, B_1\ldots B_{s-1} }$
which are parallel to $V^A$ in $s-t-1$  indices and transversal in
the rest ones. Let some solution of
(\ref{Rvac}) which describes the $(A)dS_d$ background be fixed.
The linearized HS curvature $R_1$ has the following
simple form
\bee
\label{R1A}
R_1^{A_1 \ldots A_{s-1}, B_1\ldots B_{s-1} } &=& D_0
(\go_1^{A_1 \ldots A_{s-1}, B_1\ldots B_{s-1}}) =
d \go_1^{A_1 \ldots A_{s-1}, B_1\ldots B_{s-1} }\nn\\
 &{}&\ls\ls\ls\ls\ls\ls\ls\ls\ls +(s-1)\Big(
\go_0^{\{A_1}{}_{C}\wedge
\go_1^{C A_2 \ldots A_{s-1}\}, B_1\ldots B_{s-1} }
+\go_0^{\{B_1}{}_{C}\wedge
\go_1^{ A_1 \ldots A_{s-1}, C B_2\ldots B_{s-1}\} }\Big )\,.
\eee
It is manifestly invariant under the linearized $HS$ gauge transformations
\be
\label{litr}
\delta \go{}^{A_1 \ldots A_{s-1}, B_1\ldots B_{s-1} }(x) =
D_0 \gvep {}^{A_1 \ldots A_{s-1}, B_1\ldots B_{s-1} } (x)
\ee
because $D_0^2 \equiv R(\go_0 )=0$.
The $(A)dS$ covariant form of the free HS action of
\cite{LV} is \cite{5d}
\bee
\label{gcovdact}
S^s_2&=&\half
\sum_{p=0}^{s-2}a (s,p)
\gep_{A_1 \ldots A_{d+1}}\int_{M^d} E_0^{A_5}\wedge\ldots
\wedge E_0^{A_{d}}V^{A_{d+1}} V_{C_1}\ldots V_{C_{2(s-2-p)}}\nn\\
&{}&\ls\ls\ls \wedge
R_1^{A_1 B_1 \ldots B_{s-2}}{}_,{}^{A_2 C_1 \ldots C_{s-2-p}
D_1\ldots D_p}\wedge R_1^{A_3}{}_{B_1 \ldots B_{s-2},}{}^{ A_4 C_{s-1-p} \ldots
C_{2(s-2-p)}}{}_{D_1\ldots  D_p }\,,
\eee
where $E^A_0 = D_0 (V^A )$.
The coefficients
\be
\label{al}
a (s,p) = \tilde {a} (s) (-\Lambda)^{-(s-p-1)}
\frac{(d-5 +2 (s-p-2))!!\, (s-p-1)}{  \,(s-p-2)!}\,
\ee
are fixed by the condition that
the action is independent of all those components of
$\go_1^{A_1 \ldots A_{s-1}, B_1\ldots B_{s-1} }$ for which
$V_{B_1}V_{B_2}\go_1^{A_1 \ldots A_{s-1}, B_1\ldots B_{s-1} }\neq 0$.
As a result of this condition, the free action
(\ref{gcovdact}) depends only on the frame type
 dynamical HS field
$e{}^{a_1 \ldots a_{s-1} }$  and the Lorentz connection type
auxiliary field $\go{}^{a_1 \ldots a_{s-1}, b }$ expressed
in terms of the first derivatives of  $e{}^{a_1 \ldots a_{s-1} }$
  by virtue of
its equation of motion equivalent to the ``zero torsion condition"
\be
\label{ctor}
0= T_1 {}_{A_1 \ldots A_{s-1} }\equiv
R_{1\,A_1 \ldots A_{s-1} }{}_{,B_1\ldots B_{s-1}} V^{B_1} \ldots
V^{B_{s-1}}\,.
\ee
Plugging  the expression for
$\go{}^{a_1 \ldots a_{s-1}, b }$ back into (\ref{gcovdact})
gives rise to the free HS action expressed entirely
(modulo total derivatives) in terms
of $e_\un{}^{a_1 \ldots a_{s-1} }$ and its first derivatives.
Since the linearized curvatures (\ref{R1A})
are invariant under the Abelian HS gauge
transformations (\ref{litr}),
the resulting action has required
HS gauge symmetries and therefore describes
 the free field HS dynamics in $(A)dS_d$.
In particular, the generalized Lorentz type transformations
with the gauge parameter (\ref{ltr})
guarantee that only the totally symmetric part of the gauge
field  $e_\un{}^{a_1 \ldots a_{s-1} }$
equivalent to $\varphi_{m_1\ldots m_s}$ contributes to the action.
Although this action is defined to be independent of the ``extra fields''
$ \go{}^{a_1 \ldots a_{s-1}, b_1\ldots b_t }$ with $t\geq 2$,
one has to express the extra fields in terms of
the dynamical HS fields because they contribute beyond the linearized
approximation.
The fields $ \go{}^{a_1 \ldots a_{s-1}, b_1\ldots b_t }$
 with $t>0$    express via
up to order $t$ derivatives of the dynamical field by virtue of
certain constraints \cite{Fort1,LV} analogous to the zero torsion
condition in gravity. As a result, the condition that the free action
is independent of the extra fields is equivalent to the condition that
it does not contain higher derivatives.

\section{$(A)dS_d$ bosonic higher spin algebra}
\label{Oscillator Realization}

{}From the analysis of  section \ref{Free Fields} it is clear that,
to reproduce  the correct set of HS gauge fields, one has to find such
an algebra $g$ which  contains $h=o(d-1,2)$ or  $h=o(d,1)$
 as a subalgebra and decomposes under the adjoint action
of $h$ in $g$ into a sum of
irreducible finite-dimensional modules over $h$
 described by various two-row rectangular
traceless Young tableaux.
Such an algebra was described recently by Eastwood in \cite{East}
as the algebra of conformal HS symmetries of
the free  massless Klein-Gordon equation in $d-1$ dimensions.
Here we give a slightly different definition of this algebra
which is more suitable for the analysis of the HS interactions.

Consider oscillators $Y_i^A$ with $i=1,2$ satisfying the
commutation relations
\be
\label{defr}
[Y_i^A , Y_j^B ]_* = \gvep_{ij}\eta^{AB}\,,\qquad \gvep_{ij}= -
\gvep_{ji}\,,\quad \gvep_{12}=1\,,
\ee
where  $\eta^{AB}$ is the invariant symmetric  form of
$o(n,m)$. For example, one can interpret these
oscillators as conjugated coordinates and momenta $Y^A_1 = P^A$,
$Y^B_2 = Y^B$.
$\eta_{AB}$ and $\gvep^{ij}$ will be used to raise and
lower indices in the usual manner
$
A^A = \eta^{AB} A_B
$,
$
a^i =\gvep^{ij}a_j
$,
$
a_i =a^j \gvep_{ji}\,.
$

We use the Weyl (Moyal) star product
\be
\label{wstar}
(f*g)(Y) =\f{1}{\pi^{2(d+1)}}
\int dS dT f(Y+S) g(Y+T)\exp -2 S^A_i T_A^i\,.
\ee
$[f ,g ]_* = f*g - g*f$, $\{f,g\}_* = f*g + g*f$.
The associative algebra of polynomials
with the $*$ product law generated via
(\ref{defr}) is called Weyl algebra $A_{n+m}$. Its generic
element is
$f(Y) = \sum \phi^{i_1 \ldots i_n}_{A_1 \ldots A_n} Y_{i_1}^{A_1}\ldots
Y_{i_n}^{A_n}\,$
or, equivalently,
\be
\label{exp}
f(Y) = \sum_{m,n} f_{A_1 \ldots A_m\,,B_1 \ldots B_n} Y_{1}^{A_1}\ldots
Y_{1}^{A_m}Y_{2}^{B_1}\ldots
Y_{2}^{B_n}\,
\ee
with the coefficients $f_{A_1 \ldots A_m\,,B_1 \ldots B_n}$
symmetric in the indices $A_i$ and $B_j$.

Various bilinears built from the oscillators $Y_i^A$ form the
Lie algebra $sp(2(n+m))$ with respect to star commutators. It contains
the subalgebra $o(n,m)\oplus sp(2)$ spanned by the
mutually commuting generators
\be
\label{t}
T^{AB} = -T^{BA} =\half  Y^{iA} Y^B_i\,,\qquad
t_{ij} =t_{ji} = Y^A_i Y_{jA}  \,.
\ee
Consider the subalgebra
$S$ spanned by the $sp(2)$ singlets $f(Y)$
\be
\label{sp2}
[t_{ij} , f(Y) ]_* =0\,.
\ee
Eq.(\ref{sp2}) is equivalent to
$
\Big(Y^{Ai} \f{\p}{Y^A_j}  + Y^{Aj} \f{\p}{Y^A_i}
\Big) f(Y) =0\,.
$
For the expansion (\ref{exp})  this condition implies
that the coefficients
$f_{A_1 \ldots A_m\,,B_1 \ldots  B_n}$ are nonzero
only if $n=m$
and that symmetrization over any $m+1$ indices of
$f_{A_1 \ldots A_m\,,B_1 \ldots  B_m}$ gives zero, i.e. they
 have the symmetry properties of a two-row rectangular
Young tableau. As a result, the gauge fields of $S$ are
\be
\label{gaug}
\go(Y|x) = \sum_{l=0}^\infty \go_{A_1 \ldots A_l\,,B_1 \ldots B_l}(x)
 Y_{1}^{A_1}\ldots Y_{1}^{A_l}Y_{2}^{B_1}\ldots Y_{2}^{B_l}
\ee
with the component gauge fields $\go_{A_1 \ldots A_l\,,B_1 \ldots
B_l}(x)$ taking values in all two-row rectangular Young tableaux
of $o(n+m)$.

The algebra $S$ is not simple.
It contains the two-sided ideal $I$ spanned by the elements
of the form
\be
\label{I}
g=t_{ij}*g^{ij}\,,
\ee
where $g^{ij}$ transforms as a symmetric tensor with respect to
$sp(2)$, i.e.,
$
[t_{ij}\,,g^{kl}]_* = \delta_i^k g_j{}^l +\delta_j^k g_i{}^l
+\delta_i^l g_j{}^k +\delta_j^l g_i{}^k
$.
(Note that $t_{ij}*g^{ij} =
g^{ij} *t_{ij}$.)
Actually, from (\ref{sp2}) it follows that
$f*g,\, g*f \in I$\, $\forall f\in S$, $g\in I$.
  Due to the definition (\ref{t}) of $t_{ij}$,
the ideal $I$ contains all traces
of the two-row Young tableaux. As a result, the algebra
$S/I$ has only traceless two-row
tableaux in the expansion (\ref{gaug}). (Let us note that
this factorization is not optional because
some of the traces
of  two-row rectangular tableaux are not themselves
two-row rectangular tableaux and may not admit
a straightforward interpretation in terms of HS fields.)
The algebra  $S/I$ was identified by Eastwood in \cite{East}
as conformal HS algebra in $d-1$ dimensions.

For the complex algebra $S/I$ we will use
notation $hgl(1/sp(2)[n+m]|{\bf C})$.
Its real form corresponding to a unitary HS theory in the
$AdS$ case of $n=2$ will be called
 $hu(1/sp(2)[n,m])$. The meaning of this notation is
as follows. According to \cite{KV},
$hgl(p,q|2r)$ is the superalgebra of $(p+q)\times (p+q)$ matrices
whose elements are arbitrary even (odd) power polynomials
of $2r$ pairs of oscillators in the diagonal (off-diagonal)
blocks. Because of the $sp(2)$ invariance condition (\ref{sp2}),
in our case only even functions of oscillators appear.  So we discard
the label $q$ in the notation $hgl(p,q|2r)$.
The label $sp(2)[n,m]$
means that the appropriate
quotient of the centralizer in
$hgl(1|2(n+m))$ with respect to the $sp(2)$ subalgebra, which
commutes with the  $o(n,m)$ spanned by bilinears of
oscillators, is taken.  $o(n,m)$
is  the subalgebra of  $hu(1/sp(2)[n,m])$.

Note that the described construction  of the HS algebra
is analogous to that of the $AdS_7$ HS algebra given by
Sezgin and Sundell in  \cite{7}
 in terms of spinor oscillators with the symmetric $7d$
charge conjugation matrix in place of the metric tensor in
(\ref{defr}). Also let us note that the key role of
the algebra $sp(2)$ in the analysis of HS dynamics explained below
is reminiscent of the role of $sp(2)$ in the two-time
approach developed by Bars \cite{Bars}.  In \cite{BR},
the $sp(2)$ invariant technics was applied to the description of
interacting massless fields. The important difference
 is that in our case the $sp(2)$ invariance condition
acts in the fiber space described by polynomials of the auxiliary
variables $Y^A_i$, reducing the set of fields
appropriately, while the  $sp(2)$ algebra in the models of \cite{Bars,BR}
acts  on the base.

\section{Twisted adjoint representation and\\ Central On-Mass-Shell theorem}
\label{Twisted representation and Central On-Mass-Shell Theorem}

Now we are in a position to define the twisted adjoint
representation which describes  the
HS Weyl 0-forms.
Let a  HS algebra admit such an involutive automorphism $\tau$
(i.e., $\tau(f*g) = \tau(f)*\tau (g)$, $\tau^2 =1$) that its action on the
elements of the $(A)dS_d$ subalgebra is
\be
\tau (P^a ) = -P^a\,,\qquad
\tau (L^{ab} ) = L^{ab}\,.
\ee
Once the Lorentz algebra is singled out by the compensator, the
automorphism $\tau$ describes the reflection with respect to the
compensator vector. In particular, for the HS algebra under investigation
we set
$
\tau (Y^A_i) = \tilde{Y}^A_i\,,
$
where
\be
\label{tila}
\tilde{A}^A = A^A -\f{2}{V^2}V^A V_B A^B \,.
\ee
for any  vector $A^A$.
Let us introduce notations
\be
A^{A}_i = {}^\pa A^{ A}_i + {}^\pe A^{ A}_i\,,\qquad
{}^\pa A^{ A}_i =  \f{1}{V^2} V^A V_B A^B_i\,,\quad
{}^\pe A^{ A}_i =  A^A_i - \f{1}{V^2} V^A V_B A^B_i\,.
\ee
We have ${}^\pa \tilde{A}^A_i=-{}^\pa A^A_i$ and
${}^\pe \tilde{A}^A_i={}^\pe A^A_i$.
The action of $\tau$ on a general element is
\be
\tau (f(Y )) = \tilde{f} (Y) \equiv f (\tilde{Y} )\,.
\ee
It is elementary to see that  $\tau$ is an
automorphism of the Weyl algebra.  Since $\tau (t_{ij}) = t_{ij}$,
the same is true for the HS algebra. To simplify analysis we will
assume that  $\tau$ is $x$-independent that requires
$V^A$ to be  $x$-independent.

Let $C(Y|x)$ be  a 0-form in the HS algebra linear space, i.e.
$[t_{ij} , C]_* =0$ with the ideal $I$ factored out.
The covariant derivative in the twisted adjoint representation is
\be
\tilde{D} (C)= d C + \go *C - C*\tilde{\go}\,,
\ee

Central On-Mass-Shell theorem  formulated in \cite{5d}
in terms of Lorentz components of $C(Y|x)$ states  that the
equations for totally symmetric free
massless fields in $(A)dS_d$ can be formulated in the form
\be
\label{CMS1}
R_1 ({}^\pa Y , {}^\pe Y ) = \half E_0^A \wedge  E_0^B
\f{\p^2}{\p Y^A_i \p Y^B_j} \gvep_{ij} C(0,{}^\pe Y)\,,
\ee
\be
\label{CMS2}
\tilde{D}_0 (C) =0\,,
\ee
where
$
R_1 (Y) = d\go(Y) + \go_0*\wedge \go +\go *\wedge \go_0\,,
$
$
\tilde{D}_0 (C)= d C + \go_0 *C - C*\tilde{\go}_0
$
and $\go_0 = \go_0^{AB} (x) T_{AB}$ where
$\go_0^{AB} (x)$  satisfies (\ref{Rvac}) to describe
the $(A)dS_d$  background.

The components of the expansion of the
0-forms $C(0,{}^\pe Y)$ on the r.h.s.
of (\ref{CMS1}) in powers of $Y^A_i$
are $V$ transversal. These are the
HS Weyl 0-forms $c^{a_1 \ldots a_{s}, b_1\ldots b_{s} }$
described by the traceless two-row rectangular Lorentz
Young tableaux of length $s$. They  parametrize
those components of the HS field strengths that remain
nonvanishing when the field equations and constraints on extra fields
are satisfied. For example, the Weyl tensor in gravity ($s=2$)
parametrizes the components of the Riemann tensor
allowed to be nonvanishing when  the zero-torsion
constraint and Einstein equations (requiring the Ricci tensor to
vanish) are imposed. The equation (\ref{CMS2}) describes the
consistency conditions for the HS equations and also dynamical
equations for spins 0 and 1. (Dynamics of a massless scalar
was described this way in \cite{SHV}.)
In addition they
express an infinite set of auxiliary fields  contained
in $C$ in terms of derivatives of the dynamical HS fields.

The key fact is that the equations (\ref{CMS1}) and (\ref{CMS2})
are consistent, i.e., the application of the covariant
derivative to the l.h.s. of (\ref{CMS1}) and (\ref{CMS2})
does not lead to new conditions. The only nontrivial point is to
check that
\be
 \gvep_{ij} D_0 \left (  E_0^A \wedge  E_0^B
\f{\p^2}{\p Y^A_i \p Y^B_j}  C(0,{}^\pe Y) \right )=0
\ee
as a consequence of (\ref{CMS2}).  It is convenient to
use the following forms of the covariant derivatives in the adjoint and
twisted adjoint representations
\be
\label{ad}
D_0 = D^L_0 -\Lambda  E^A_0 V^B \Big ( {}^\pe Y_{Ai}
\f{\p}{\p {}^\pa Y^{ B}_i} -
{}^\pa Y_{Bi}\f{\p}{\p {}^\pe Y^A_i} \Big )\,,
\ee
\be
\label{twad}
\tilde{D}_0 = D_0^L - 2\Lambda  E_0^A V^B \Big ( {}^\pe Y_{A}^i {}^\pa
Y_{Bi} -
\f{1}{4}\gvep^{ji}
\f{\p}{ \p {}^\pe Y^{ A j}\p {}^\pa Y^{Bi }} \Big )\,,
\ee
where the Lorentz covariant derivative is
$
D_0^L = d +  \go_0^{L\, AB}
{}^\pe Y_{Ai}\f{\p}{\p {}^\pe Y^B_i}  \,.
$
One observes that $D_0^L$ commutes with
 $E_0^A \wedge  E_0^B
\f{\p^2}{\p Y^A_i \p Y^B_j} \gvep_{ij}$. As a result, it remains to check
the frame dependent terms. These all vanish
either because the r.h.s. of (\ref{CMS1}) is
${}^\pa Y^A_i$--independent or because of  the identity
$
E_0^A \wedge  E_0^B \wedge  E_0^C
\f{\p^3}{\p Y^A_i \p Y^B_j \p Y^C_k} C(Y)\equiv 0\,
$
valid as a consequence of the total antisymmetrization over
the indices  $i,j$ and $k$
which take two values  (equivalently, because the $C(Y)$
does not contain  Young tableaux with more than two rows).

The adjoint covariant derivative (\ref{ad}) and twisted adjoint
covariant derivative (\ref{twad})
commute with the operators $N^{ad}$ and  $N^{tw}$, respectively,
\be
N^{ad}= Y^A_i\f{\p}{\p Y^A_i}\,,\qquad
N^{tw}= {}^\pe Y^A_i\f{\p}{\p {}^\pe Y^A_i}-
{}^\pa Y^A_i\f{\p}{\p {}^\pa Y^A_i}\,.
\ee
This means that the free field equations (\ref{CMS1}) and
(\ref{CMS2}) decompose into independent subsystems
for the sets of fields satisfying
\be
\label{spin}
N^{ad} \go = 2(s-1) \go\,, \qquad N^{tw} C = 2s C\,, \qquad s \geq 0\,.
\ee
(Note that the operator $N^{tw}$ does not have negative eigenvalues
when acting on tensors with the symmetry property of
a two-row rectangular Young tableau because
having more than a half of vector indices
aligned along  $V^A$ would imply symmetrization
over more than a half of indices,
thus giving zero.)

The set of fields
singled out by (\ref{spin}) describes a massless spin $s$.
As expected, the massless scalar field is described  only in terms of
the 0-form $C(Y|x)$, having no associated gauge field. In terms of
Lorentz irreducible components, the spin $s$ gauge connections take
values in
the representations (\ref{dia}) with various $0\leq t\leq s-1$ while
the spin $s$ Weyl tensors take values in the Lorentz representations
\begin{picture}(43,15)
\put( 9,12){\scriptsize   p}    
\put( 9,-5){\scriptsize   s}    
{\linethickness{.250mm}
\put(00,10){\line(1,0){30}}
\put(00,05){\line(1,0){30}}
\put(00,00){\line(1,0){15}}
\put(00,00){\line(0,1){10}}%
\put(05,00.0){\line(0,1){10}}
\put(10,00.0){\line(0,1){10}}
\put(15,00.0){\line(0,1){10}}
\put(20,05.0){\line(0,1){05}}
\put(25,05.0){\line(0,1){05}}
\put(30,05.0){\line(0,1){05}}
}
\end{picture} with various $p\geq s$. (Note that the missed cells
 compared to the rectangular diagram of the length
of the upper row correspond to the Lorentz
invariant direction along $V^A$.) We see that  the
twisted adjoint action of the $(A)dS_d$ algebra on the HS algebra
decomposes into an infinite set of infinite-dimensional submodules
associated with different spins, while its adjoint action
decomposes into an infinite set of finite-dimensional submodules.
This fits the fact (see e.g. \cite{sah3}), that the space of
physical states is described by
the  0-form  sector  which therefore has to form an infinite module
over a space-time symmetry for
any dynamical system with infinite number of degrees of freedom.

\section{Nonlinear equations}
\label{Nonlinear Equations}

Following to the
 standard approach in HS theory \cite{Ann}
we will  look for a nonlinear deformation of the free field
equations (\ref{CMS1}) and (\ref{CMS2}) in the form of a free
differential algebra
$dW^\alpha = F^\alpha (W)\,,$
satisfying the consistency condition
$F^\gb \f{\delta F^\ga}{\delta W^\gb}\equiv 0 \,,$
where $W^\ga$ describes all dynamical variables in the system (i.e.
all 1-forms $\go$ and 0-forms $C$ in our case).
The equations (\ref{CMS1}) and (\ref{CMS2})
acquire nonlinear corrections once the fields
$\go_0$ and $\go$ are interpreted, respectively,
as the vacuum and fluctuational parts
of some dynamical field. In this case
one has to replace the background covariant derivatives by the
full ones that gives rise to nontrivial Bianchi relations because
$D^2$ is proportional to
the generalized Weyl tensors (in particular, to the
usual Weyl tensor in the spin 2 sector).
As shown in \cite{gol} the
key principle that fixes a form of the $4d$ nonlinear
field equations found in \cite{more} is the condition
that the local Lorentz symmetry remains a proper subalgebra
of the infinite-dimensional HS algebra. This requirement
 guarantees that the nonlinear theory can be  interpreted
in terms of usual (finite-component) tensor fields. Otherwise,
a deformation of the algebra might affect the form of the
commutator of the Lorentz generators which may get admixture of
infinity of HS generators with field-dependent coefficients.
The same principle was used for the derivation of $3d$ equations
\cite{PV}.
The observation that allows us to
build bosonic HS equations in any $d$ is that in this case one has
to require the $sp(2)$ algebra, which singles out the
HS algebra spanned by two-row rectangular tensor elements, to
remain undeformed at the nonlinear level. Otherwise,
the condition (\ref{sp2}) would not allow a
meaningful extension beyond the free field level, i.e., the resulting
system may admit no interpretation in terms of the original HS
tensor fields described by two-row rectangular Young tableaux,
allowing, for example, nonlinear corrections in
sectors of tensors which are absent at the linearized level.

The formulation of nonlinear HS equations in any dimension
is nearly identical to that
of $4d$ and $3d$  HS field equations given in \cite{more,PV}
in terms of spinors.  First,  we double a number of oscillators
$Y^A_i \to (Z^A_i , Y^A_i )$. The space of functions $f(Z,Y)$ is endowed
with the star product
\be
\label{star}
(f*g)(Z,Y) =\f{1}{\pi^{2(d+1)}}
\int dS dT f(Z+S,Y+S) g(Z-T, Y+T)
\exp -2 S^A_i T_A^i \,,
\ee
which is associative, normalized so that $1*f =f*1= f$ and
gives rise to the  commutation relations
\be
\label{defr1}
[Y_i^A , Y_j^B ]_* = \gvep_{ij}\eta^{AB}\,,\qquad
[Z_i^A , Z_j^B ]_* = - \gvep_{ij}\eta^{AB}\,, \qquad
[Y_i^A , Z_j^B ]_* =0\,.
\ee
The star product (\ref{star}) describes a
normal-ordered basis in $A_{2(n+m)}$ with respect to creation
and annihilation operators $Z-Y$ and $Y+Z$, respectively.
For $Z$ independent elements (\ref{star}) coincides with (\ref{wstar}).
The following  useful formulae are true
\be
\label{uff1}
Y^A_i * = Y^A_i + \half\left ( \f{\p}{\p Y_A^i}-\f{\p}{\p Z_A^i}\right )
 \,,\qquad
Z^A_i * =
Z^A_i + \half\left ( \f{\p}{\p Y_A^i}-\f{\p}{\p Z_A^i}\right )
 \,,
\ee
\be
\label{uff2}
* Y^A_i  = Y^A_i - \half \left( \f{\p}{\p Y_A^i}+\f{\p}{\p Z_A^i}\right )\,,
\qquad
* Z^A_i  = Z^A_i + \half \left (\f{\p}{\p Y_A^i}+\f{\p}{\p Z_A^i}\right )\,.
\ee

Important property of the star product (\ref{star}) is that it
admits inner Klein operators. Indeed, it is elementary to see
with the aid of (\ref{star}) that the element
\be
{\cal K} = \exp {-2z_i y^i } \,,
\ee
where
\be
y_i = \f{1}{\sqrt{V^2}} V_B Y^B_i\,,\qquad
z_i = \f{1}{\sqrt{V^2}} V_B Z^B_i\,
\ee
has the  properties
\be
\label{K}
{\cal K} *f = \tilde{f}*  {\cal K}\,,\qquad {\cal K} *{\cal K}=1\,,
\ee
where  $\tilde{f}(Z,Y) = f(\tilde{Z} , \tilde{Y})$.
This follows from the following formulae
\be
{\cal K} *f = \exp {-2z_i y^i }
f(Z^A_i - \f{1}{V^2} V^A V_B (Z^B_i -Y^B_i ),
Y^A_i + \f{1}{V^2} V^A V_B (Z^B_i -Y^B_i ))\,,
\ee
\be
\label{fK}
f* {\cal K}  = \exp {-2z_i y^i }
f(Z^A_i - \f{1}{V^2} V^A V_B (Z^B_i +Y^B_i ),
Y^A_i - \f{1}{V^2} V^A V_B (Z^B_i +Y^B_i ))\,.
\ee

We introduce the fields $W(Z,Y|x)$, $B(Z,Y|x)$ and   $S(Z,Y|x)$,
where $B(Z,Y|x)$ is a 0-form while $W(Z,Y|x)$ and $S(Z,Y|x)$ are
connection 1-forms in space-time and auxiliary $Z^A_i$ directions,
respectively
\be
W(Z,Y|x)=dx^\un W_\un (Z,Y|x)\,,\qquad
S(Z,Y|x)=dZ^A_i S^i_A (Z,Y|x)\,.
\ee
The fields $\go$ and $C$ are identified with the ``initial data''
for the evolution in $Z$ variables
\be
\go (Y|x) = W(0,Y|x)\,,\qquad C (Y|x) = B(0,Y|x)\,.
\ee
The $Z$ - connection $S$ will be determined modulo gauge ambiguity in terms
of $B$.  The differentials satisfy the standard anticommutation
relations
$
dx^\un dx^\um = - dx^\um dx^\un\,,
$
$
dZ^A_i dZ^B_j = - dZ^B_j dZ^A_i\,,
$
$
dx^\un dZ^B_j = - dZ^B_j dx^\un
$
and commute to all other variables (from now on we discard the wedge
symbol).

The full nonlinear system of HS equations is
\be
\label{WW}
dW+W*W =0\,,
\ee
\be
\label{DS}
dS+W*S+ S*W =0 \,,
\ee
\be
\label{DB}
dB+W*B-B*\tilde{W} =0\,,
\ee
\be
\label{SB}
S*B = B*\tilde{S}\,,
\ee
\be
\label{SS}
S*S = -\half( dZ^A_i dZ_A^i  + 4\Lambda^{-1} dz_i dz^{i} B*\K )\,,
\ee
where $\tilde{S}(dZ , Z,Y) = S(\tilde{dZ} , \tilde{Z},\tilde{Y})$
and $dz_i = \f{1}{\sqrt{V^2}} V_B dZ^B_i$.
In terms of  a noncommutative connection
$
\W = d +W +S
$
 the system (\ref{WW})-(\ref{SS})
reads
\be
\W *\W =  -\half( dZ^A_i dZ_A^i  + 4\Lambda^{-1}  dz_i dz^{i} B*\K )\,,\qquad
\W * B = B *\tilde{\W}\,.
\ee
We see that   $ dz_i dz^{i} B*\K$ is
the only nonzero component of the  curvature.
Note that $B$ has dimension $cm^{-2}$ to match the
Central On-Mas-Shell theorem (\ref{CMS1}) upon identification
of $B$ with $C$ in the lowest order. So,
$\Lambda^{-1}  B$ is dimensionless. Since the $B$ dependent
part of the equation (\ref{SS}) is responsible for interactions,
this indicates that taking the flat limit
may be difficult in the interacting theory if there is no other
contributions to the cosmological constant  due to some
condensates which break  the HS gauge symmetries.

The system is formally consistent in the sense that the associativity
relations $\W*(\W*\W)=(\W*\W )*\W$ and $(\W*\W)*B = B* (\W*\W )$,
equivalent to Bianchi identities,
are respected by the equations  (\ref{WW})-(\ref{SS}).
The only nontrivial part of this property might be that for the
relationship $(S*S)*S=S*(S*S)$ in the sector of $(dz_i )^3$ due to the
second term on the r.h.s. of (\ref{SS}) since
$B*\K$ commutes with everything except  $dz_i $ to which it
anticommutes. However, this does not break the consistency of the system
because $(dz_i)^3\equiv 0$.
As a result, the equations (\ref{WW})-(\ref{SS}) are consistent
as ``differential''
 equations with respect to $x$ and $Z$ variables.
A related statement is that the equations (\ref{WW})-(\ref{SS})
are invariant under the gauge transformations
\be
\label{gxz}
\delta \W = [\gvep , \W ]_* \,,\qquad \delta B = \gvep * B -
B*\tilde{\gvep}
\ee
with an arbitrary gauge parameter $\gvep (Z,Y|x)$.

Let us show that the condition (\ref{sp2}) admits a proper deformation
to the full nonlinear theory, i.e. that
there is a proper deformation
$t^{int}_{ij}$ of  $t_{ij}$ that allows us to impose the
conditions
\be
\label{tonlc}
D(t^{int}_{ij})=0\,,\qquad
 [S, t^{int}_{ij}]_*=0\,,\qquad
B * \tilde{t}^{int}_{ij} - t^{int}_{ij} *B =0\,,
\ee
which amount to the original conditions $[t_{ij}, W]_* =[t_{ij}, B]_*=0 $
 in the free field limit. Indeed, let us introduce
the  generators of the diagonal $sp^{tot}(2)$
algebra
\be
\label{Tzy}
t^{tot}_{ij} =  Y^A_i Y_{Aj}  - Z^A_i Z_{jA}  \,.
\ee
 In any $sp(2)$ covariant gauge (in which $S^i_A$
is expressed in terms of $B$ with no external $sp(2)$
noninvariant parameters - see below)
$sp^{tot}(2)$ acts on $S^i_A$ as
$
[t^{tot}_{ij} \,, S^A_n ]_* =  \gvep_{in}S^A_{j}+ \gvep_{jn}S^A_{i}\,
$
provided that $B$ is a $sp(2)$ singlet.

Setting $S= dZ_i^A S^i_A$, the equation (\ref{SS}) gets the form
$
[S^i_A \,,S^j_B ]_* = -\gvep^{ij} (\eta_{AB} -4 V_A V_B B*\K )\,.
$
{}From (\ref{SB}) and (\ref{K}) it follows  that
\be
S^i_A *B*\K = B*\K *( S^i_A - \f{2}{V^2}V^A V^B S^i_B )\,.
\ee
As a result, we get
\be
\label{sb}
[s^i , s^j ]_* = -\gvep^{ij} (1 + 4\Lambda^{-1} B*\K)\,,\qquad
s^i *B*\K = - B*\K*s^i\,,\qquad s^i =  \f{1}{\sqrt{V^2}}V^A S^i_A \,
\ee
and $[{}^\pe S^i_A, s^j]_* =0$,
$[{}^\pe S^i_A, {}^\pe S_B^j]_* =-\gvep^{ij}{}^\pe \eta_{AB} $.
The commutation relations (\ref{sb}) have a form of the
deformed oscillator algebra \cite{Aq}
originally found by Wigner in \cite{wig} in a particular representation.
Its key property is that the operators
$
{\cal T}_{ij} = - \half \{S^A_i\,, S_{Aj} \}_*\,
$
form $sp(2)$ and
$
[\T_{ij} \,, S^A_n ]_* =  \gvep_{in}S^A_{j}+ \gvep_{jn}S^A_{i}\,.
$

As a result,  the operators
\be
t^{int}_{ij}=t^{tot}_{ij}-\T_{ij}\,
\ee
form  $sp(2)$ and commute with $S^A_i$.
The conditions
(\ref{tonlc}) are equivalent to the usual $sp(2)$ invariance conditions
for the fields $W$ and $B$ and are identically satisfied
 on $S_i^A$, which essentially means that nonlinear corrections
due to the evolution along $Z$-directions in the noncommutative space
do not affect $sp(2)$.
In the free field limit with $S^A_i = Z^A_i$, $t^{int}_{ij}$ coincides
with (\ref{t}).
One uses  $t^{int}_{ij}$ in the nonlinear
model the same way as  $t_{ij}$ (\ref{t}) in the
free one to impose (\ref{tonlc}) and to factor out
the respective ideal $I^{int}$.
This guarantees that the nonlinear field
equations described by Eqs.(\ref{WW})-(\ref{SS})
and (\ref{tonlc}) make sense for the
HS fields associated with  $hu(1/sp(2)[n,m])$.

\section{Perturbative analysis}

The perturbative analysis of the equations (\ref{WW})-(\ref{SS})
is analogous to that carried out in
spinor notations in \cite{more} for the $4d$
case. Let us set
\be
W=W_0 +W_1 \,,\qquad S= S_0 +S_1 \,,\qquad B=B_0 +B_1
\ee
with the vacuum solution
\be
\label{BS0}
B_0 = 0 \,,\qquad S_0 = dZ^A_i Z_A^i \,,\qquad
W_0 =  \half \go_0^{AB} (x)  Y^i_A Y_{iB}\,,
\ee
where $\go_0^{AB} (x)$ satisfies the zero curvature conditions to describe
$(A)dS_d$. {}From (\ref{SB}) and (\ref{BS0}) we get
\be
\label{BC}
B_1 = C(Y|x)\,.
\ee

Let us consider the equation (\ref{SS}).
 First of all we observe that using
the gauge ambiguity (\ref{gxz})
 we can set all components of ${}^\pe S^A_{1i}$
equal to zero, i.e. to set
$
S_1 = dz_i s_1^i (z,Y|x) \,.
$
The leftover gauge symmetry parameters are  ${}^\pe Z$
independent.
The equations (\ref{DS}) and (\ref{SB}) then require
the fields $W$ and $B$ also to be ${}^\pe Z$ independent.
So, the dependence on $Z$ now enters only through $z_i$.
As a result, (\ref{SS}) amounts to the first equation in (\ref{sb}).
With the help of (\ref{fK}) one obtains in the first order
\be
\label{partS}
\p^i s^j_1 - \p^j  s^i_1
=-4\Lambda^{-1} \gvep^{ij}
C(-{}^\pa Z,{}^\pe Y)\exp{-2 z_k y^k }\,,
\ee
where $\p^i = \f{\partial}{\p z_i}$
The generic solution of this equation is
\be
\label{S1}
s^j_1 =\p^j  \varepsilon_1 +2\Lambda^{-1} z^j
\int^1_0dt\,t
C(-t{}^\pa Z,{}^\pe Y )\exp{-2t z_i y^i }\,.
\ee

The ambiguity in the function $ \varepsilon_1=\varepsilon_1 (Z,Y|x) $
manifests invariance under the gauge transformations
(\ref{gxz}).
It is convenient to fix a gauge by requiring $ \partial^i
\varepsilon_1=0 $ in (\ref{S1}). (This  gauge is  covariant because
it  involves no external parameters carrying nontrivial
representations of $sp(2)$.)
This gauge fixing is not complete as
it does not fix the gauge transformations with $Z$ independent parameters
 \be
\label{WE}
\varepsilon_1 (Z,Y|x)=\varepsilon_1 (Y|x) \,.
\ee
 As a result, the field $ S $ is
expressed in terms of $ B $.  It is not surprising
of course that the noncommutative gauge connection $S$ is
reconstructed in terms of the noncommutative curvature $B$
modulo gauge transformations.
The leftover gauge transformations
with the
parameter (\ref{WE}) identify with the HS gauge transformations
acting on the  physical HS fields.

Now, let us analyze the  equation (\ref{DS}).
In the first order, one gets
\be
\label{partw}
\p^i W_1 = d s^i_1 +W_0 * s^i_1 -  s^i_1 *W_0  \,.
\ee
Using  that generic solution of the equation
$
 \frac{\partial}{\partial z_i } \varphi(z)=\chi^i (z)
$
has the form
$
 \varphi (z)=  const +\int^1_0 dt\,z_i \, \chi^i (tz)
$
provided that
 $  \frac{\partial}{\partial z^i } \chi^i (z)
 \equiv 0 $ and $ i =1,2,$ one finds
\be
\label{W1}
W_1(Z,Y)=\go (Y) -Z^j_A V^A
 \int^1_0 dt\,(1-t)e^{-2tz_i y^i}
   E^B\f{\p}{\p Y^{jB}} C(-t{}^\pa Z, Y^\pe )\,
\ee
(note that the terms with $ z_i dS^{1\,i}$
  vanish because $ z^i z_i  \equiv 0 $). Since, perturbatively,
the system as a whole is  a consistent
 system of differential equations with respect to $
\frac{\partial}{\partial z} $  and $\frac{\partial}{\partial x} $
it is enough to analyze
the equations (\ref{WW}) and (\ref{DB}) at $ Z =0 $.
Thus, to derive dynamical HS equations,
it remains to insert (\ref{W1}) into (\ref{WW}) and (\ref{BC})
into (\ref{DB}),
interpreting $ \go (Y|x) $ and $ C(Y|x) $  as generating
functions for the HS fields. The elementary analysis of
(\ref{WW}) at $ Z=0 $ with the help of (\ref{uff1}) and (\ref{uff2})
gives (\ref{CMS1}).
For $B=C$, the equation  (\ref{DB})
amounts to  (\ref{CMS2}) in the lowest order.
Thus it is shown that the linearized part of the HS equations
(\ref{WW})-(\ref{SS}) reproduces the Central On-Mass-Shell theorem
for symmetric massless fields.
The system (\ref{WW})-(\ref{SS}) allows one to
derive systematically all higher-order corrections to the free
equations.

\section{Discussion}

The system of
gauge invariant  nonlinear dynamical equations  for totally symmetric
massless fields of all spins in $AdS_d$
presented in this paper can be  generalized to a class of models with
the  Yang-Mills groups $U(p)$, $USp(p)$ or $O(p)$.
This results from the observation that, analogously to the case of $d=4$
\cite{Ann,KV}, the system (\ref{WW})-(\ref{SS}) remains consistent
for matrix valued  fields
$W\to W_\ga{}^\gb $, $S\to S_\ga{}^\gb$ and $B\to B_\ga{}^\gb$,
$\ga,\gb = 1\ldots p$.
Upon imposing the reality conditions
\be
W^\dagger (Z,Y|x)= - W (-iZ,iY|x)\,,\qquad
S^\dagger (Z,Y|x)= - S (-iZ,iY|x)\,,
\ee
\be
B^\dagger (Z,Y|x)= - \tilde{B} (-iZ,iY|x)
\ee
this gives rise to a system with the global HS symmetry algebra
$hu(p/sp(2)[n,m])$. Here all fields, including the spin 1
fields which correspond  to
the $Z,Y$-independent part of $W_\ga{}^\gb(Z,Y|x)$, take values in
$u(p)$ which is the Yang-Mills algebra of the theory.

Combining the
antiautomorphism of the star product algebra
$\rho (f(Z,Y)) = f (-iZ,iY)$ with some antiautomorphism of the matrix
algebra generated by a nondegenerate form $\rho_{\ga\gb}$
one can impose the conditions
\be
W_\ga{}^\gb  (Z,Y|x)=
- \rho^{\gb\gga}\rho_{\delta\ga} W_\gga{}^\delta (-iZ,iY|x)\,,\quad
S_\ga{}^\gb  (Z,Y|x)=
- \rho^{\gb\gga}\rho_{\delta\ga} S_\gga{}^\delta (-iZ,iY|x)\,,
\ee
\be
B_\ga{}^\gb  (Z,Y|x)=
- \rho^{\gb\gga}\rho_{\delta\ga} \tilde{B}_\gga{}^\delta (-iZ,iY|x)\,,
\ee
which
truncate the original system to the one with the Yang-Mills
gauge group $USp(p)$ or $O(p)$ depending on whether the form
$\rho_{\ga\gb}$ is antisymmetric or symmetric, respectively.
The corresponding global HS symmetry algebras are called
$husp(p/sp(2)[n,m])$ and $ho(p/sp(2)[n,m])$,
respectively.
In this case  all fields of odd spins take values in the
adjoint representation of the
Yang-Mills group while fields of even spins take values in the opposite
symmetry second rank representation (i.e., symmetric for
 $O(p)$ and antisymmetric for
 $USp(p)$) which contains singlet. The graviton
is always the singlet spin 2 particle in the theory.
Color spin 2 particles are also included for general $p$
however.\footnote{Let us note that this does not
contradict to the no-go results of \cite{CW,ng2} because
the theory under consideration
does not allow a flat limit with  unbroken
HS symmetries and color spin 2 symmetries.} The
minimal HS theory is  based on the algebra $ho(1/sp(2)[n,m])$.
It describes  even spin particles, each in one copy. (Odd spins
do not appear because the adjoint representation of $o(1)$ is trivial.)

All HS models have essentially one
dimensionless coupling constant $g^2=|\Lambda|^{\f{d-2}{2}} \kappa^2$,
where $\kappa^2$ is the  gravitational constant.
At the level of field equations it can be rescaled away by a field
redefinition. The Yang-Mills  coupling
constant is $g^2_{YM} = |\Lambda| \kappa^2$. Unlikely to
the $4d$ case of \cite{more},
for general $d$  we do not see a  way for a consistent nontrivial
modification of the system  (\ref{WW})-(\ref{SS})  that cannot
be compensated by a  field redefinition.
This  probably means that the interaction
ambiguity in one function
found in \cite{more} is due to the possibility to treat
independently the selfdual and anti-selfdual parts of the HS curvatures,
that makes no sense for totally symmetric fields beyond $d=4$.

The results of this paper  indicate that there
may exist a large class of consistent HS theories in  $(A)dS_d$
with different spin spectra.
An important problem for the future is therefore
to extend the obtained results
beyond the class of totally symmetric gauge fields. A formulation
of arbitrary symmetry free massless fields in the flat space-time
is fairly well understood by now
(see e.g., \cite{mixed}). The general covariant
formulation in $(A)dS_d$
is still lacking however, although a number of important contributions
has been made.
In particular, the light cone formulation of
the equations of motion of generic massless fields in $AdS_d$
\cite{meq} and actions for generic massless
fields in $AdS_5$ \cite{met} were constructed by Metsaev.
In \cite{BS} an
approach to covariant description of an arbitrary representation
of $AdS_d$ algebra $o(d-1,2)$ was developed in the framework of the
radial reduction technique. The unfolded formulation
of a $5d$ HS theory with mixed symmetry fields was studied in
\cite{SS2} in the sector of  Weyl 0-forms.
The phenomenon of partial masslessness of HS fields in $(A)dS$
is another specificity of the dynamics in $(A)dS_d$ \cite{DW} .
Extension of flat space results
to $(A)dS_d$ is not  straightforward for mixed symmetry
fields because  irreducible
massless systems in $AdS_d$ reduce in the flat limit to a collection
of independent flat space massless fields \cite{BMV}.

\section*{Acknowledgments}
The author is grateful to Lars Brink for  hospitality at
the Chalmers University of Technology where a part of this work was done.
This research was supported in part by
INTAS, Grant No.00-01-254,
 and the RFBR, Grant No.02-02-17067.

\end{document}